\documentclass[sigconf]{acmart}
\AtBeginDocument{%
  \providecommand\BibTeX{{%
    \normalfont B\kern-0.5em{\scshape i\kern-0.25em b}\kern-0.8em\TeX}}}

\setcopyright{iw3c2w3}
\copyrightyear{2021}
\acmYear{2021}
\acmConference[WWW '21]{Proceedings of the Web Conference 2021}{April 19--23, 2021}{Ljubljana, Slovenia}
\acmBooktitle{Proceedings of the Web Conference 2021 (WWW '21), April 19--23, 2021, Ljubljana, Slovenia}
\acmPrice{}
\acmDOI{10.1145/3442381.3449986}
\acmISBN{978-1-4503-8312-7/21/04}

\usepackage{url}
\usepackage{booktabs}
\usepackage{tikz}
\usepackage{makecell}
\usepackage{threeparttable}
\usepackage{longtable}
\usepackage{multirow}
\usepackage{tabularx}
\usepackage{tabulary}
\usepackage{tablefootnote}

\usepackage{amssymb}
\usepackage{epstopdf}
\usepackage{graphicx}
\usepackage{subfig}
\usepackage{caption}
\usepackage{color}
\usepackage{mathtools}
\usepackage{hyperref}
\usepackage{bm}
\usepackage{dsfont}
\usepackage{enumitem}
\begin{document}
\title{Interest-aware Message-Passing GCN for Recommendation}

\author{Fan Liu$^\dag$, Zhiyong Cheng$^{\S*}$, Lei Zhu$^\ddag$, Zan Gao$^\S$, Liqiang Nie$^{\dag*}$}
\affiliation{\institution{$^\dag$School of Computer Science and Technology, Shandong University}\country{}}
\affiliation{\institution{$^\S$Shandong Artificial Intelligence Institute, Qilu University of Technology (Shandong Academy of Sciences)}\country{}}
\affiliation{\institution{$^\ddag$School of Information Science and Engineering, Shandong Normal University}\country{}}
\email{{liufancs, jason.zy.cheng, nieliqiang }@gmail.com}
\thanks{* Corresponding author: Zhiyong Cheng and Liqiang Nie}

\begin{abstract}
Graph Convolution Networks (GCNs) manifest great potential in recommendation. This is attributed to their capability on learning good user and item embeddings by exploiting the collaborative signals from the  high-order neighbors. Like other GCN models, the GCN based recommendation models also suffer from the notorious over-smoothing problem – when stacking more layers, node embeddings become more similar and eventually indistinguishable, resulted in performance degradation. The recently proposed LightGCN and LR-GCN alleviate this problem to some extent, however, we argue that they overlook an important factor for the over-smoothing problem in recommendation, that is, high-order neighboring users with no common interests of a user can be also involved in the user's embedding learning in the graph convolution operation. As a result, the multi-layer graph convolution will make users with dissimilar interests have similar embeddings. In this paper, we propose a novel Interest-aware Message-Passing GCN (IMP-GCN) recommendation model, which performs high-order graph convolution inside subgraphs. The subgraph consists of users with similar interests and their interacted items.  To form the subgraphs, we design an unsupervised subgraph generation module, which can effectively identify users with common interests by exploiting both user feature and graph structure. To this end, our model can avoid propagating negative information from high-order neighbors into embedding learning. Experimental results on three large-scale benchmark datasets show that our model can gain performance improvement by stacking more layers and outperform the state-of-the-art GCN-based recommendation models significantly. 
\end{abstract}


\maketitle
\section{Introduction}
Recommendation system has become one of the most important techniques for various online platforms. It can not only provide personalized information for an specific user from overwhelming information, but also increase the revenue for service providers. Among them, Collaborative filtering (CF) based models~\cite{netflix,he2017neural,wu2016cdae,xue2017deep,Liu2019MAML} have made substantial progress in learning user and item representations by modeling historical user-item interactions. For example, matrix factorization (MF) can directly embed user/item as a feature vector and model the user-item interactions with inner product~\cite{netflix}. Neural collaborative filtering models replace the MF interaction function of inner product with nonlinear neural networks to learn better user and item representations~\cite{he2017neural}. 

Recently, GCN-based models~\cite{berg2019gcmc,wang2019ngcf,wang2019kdd,He2020lightgcn,liu2020A2GCN} have achieved great success in recommendation due to the powerful capability on representation learning from non-Euclidean structure. The core of GCN-based models is to iteratively aggregate feature information from local graph neighbors. It has been proved to be an efficient way to distill additional information from graph structure, and thus improves user and item representation learning and alleviates the sparse problem. For example, NGCF~\cite{wang2019ngcf} has proved that exploiting high-order connectivity can help alleviate the sparsity problem in recommendation. However, it is also well-recognized that GCNs suffer from the \textit{over-smoothing} problem~\cite{wang2019ngcf}, because the graph convolution operation is actually a special kind of graph Laplacian smoothing~\cite{wang2019ngcf}, making node representations become indistinguishable after multi-layer graph convolution~\cite{Xin2020GraphHN}. As a result, most current GCN based models obtain their peak performance by stacking only few layers (e.g., 2 or 3 layers), and continuing increasing the depth will lead to sharp performance degradation. In the domain of recommendation, Chen et al.~\cite{ChenWHZW20} have empirically demonstrated that the user/item embeddings become more similar when stacking more layers in NGCF due to the over-smoothing effect. In other words, the preferences of different users become homogeneous, resulted in performance degradation in recommendation. Based on the observations, they proposed a LR-GCN model, which removes the non-linearities in GCNs to simply the network structure and introduced a residual network structure to alleviate the over-smoothing problem, achieving substantially improvement over NGCF on recommendation accuracy.

It is worth mentioning that the LightGCN proposed by He et al.~\cite{He2020lightgcn} has a similar formulation as LR-GCN. With careful experimental studies, He et al. pointed out that the feature transformation and nonlinear activation have no positive effect (or even negative effect due to the increase of training difficult) to the final performance. Therefore, they only keep the neighborhood aggregation in the LightGCN for collaborative filtering. Comparing to LR-GCN, LightGCN further removes the ``self-loop" in the aggregation operation. Although LightGCN is not dedicatedly designed for tacking the over-smoothing problem, it has almost the same formulation as LR-GCN and thus can also alleviate the over-smoothing problem to some extent. In fact, both LR-GCN and LightGCN are consistent with the recent theories in simplifying GCNs~\cite{pmlr-v97-wu19e} and can obtain the best performance with a deeper structure (e.g., 4 layers). Despite the two success GCN based models are designed for recommendation, we argue that they still design the model from the perspective of graph convolution, while have not well considered the over-smoothing problem in the domain of recommendation. 

The GCN based recommendation model is built upon a user-item graph, in which the user and item are linked according to the historical user-item interactions. The user embedding is learned by iteratively aggregating messages passed from the neighboring (both user and item) nodes. Note that the passed messages are distilled from the embeddings of neighboring nodes. When stacking $k$ layers, the information from the $k$-order neighbors, which are indirectly connected via items and users, are also involved in the embedding learning of a target node. An underlying assumption is that the collaborative signals from high-order neighbors are beneficial to the embedding learning. However, not all the information from high-order neighbors are positive in reality. In the user-item interaction graph, the high-order neighboring users could have no common or even contradictory  interest with a target user. This is highly possible, especially when the graph is constructed based on implicit feedbacks (e.g., click). In fact, the implicit feedback is more widely used over the explicit feedbacks in modern recommendation systems. The core idea behind collaborative filtering is that similar users like similar items. Therefore, the collaborative signals that we would like to exploit should be from similar users (i.e., users with similar interests). However, existing GCN-based recommendation models have not distinguished the high-order neighbors, and just simply aggregate the messages from all those neighbors to update user embeddings. As a result, the embeddings of dissimilar users are also involved in the embedding learning of a target user, negatively affecting the performance. This is also a reason of the over-smoothing effect in the GCN-based recommendation models – making the embeddings of dissimilar users to be similar. 

Motivated by the above considerations, in this paper, we propose a novel Interest-aware Messaging-Passing GCN (IMP-GCN) recommendation model, which groups users and their interacted items into different subgraphs and operates high-order graph convolutions inside subgraphs.  More specific, we adopt the simplified network structure of LightGCN, as its effectiveness has been well demonstrated in~\cite{He2020lightgcn} and it can alleviate the over-smoothing problem to some extent. The first-order graph convolution is the same as that of LightGCN. For the high-order graph convolution, only the messages from nodes in the same \textit{subgraph} are exploited to learn the node embeddings. The subgraph is generated by a proposed graph generation module, which integrates users features and graph structure to identify users with similar interests, and then constructs the subgraphs by retaining those users and their interacted items. To this end, our model can filter out the negative information propagation in the high-order graph convolution operations for the embedding learning, and thus can keep the uniqueness of users by stacking more graph convolution layers.  Extensive experiments have been conducted on three large-scale real-world datasets to validate the effectiveness of our model. Results show that our model outperforms the state-of-the-art methods by a large margin and can obtain better performance with more layers (till 7 layers) \footnote{In experiments, we found that by stacking 7 layers, a user node almost reaches all the other users in three different datasets. Therefore, no more gain after stacking 7 layers.}. This indicates that our model can benefits from higher-order neighbors by excluding negative nodes. Besides, with deep analysis on the results, we found that the negative information in the embedding propagation is the major reason for the performance degradation of existing GCN-based recommendation models in deep structure. We released the codes and involved parameter settings to facilitate others to repeat this work~\footnote{https://github.com/liufancs/IMP\_GCN.}.

In summary, the main contributes of this work are as follows:
\begin{itemize}[leftmargin=*]
	\item We step into the over-smoothing problem in existing GCN-based recommendation models and point out an overlooked factor: exploiting high-order neighbors indiscriminately makes the embeddings of users with dissimilar interests to be similar.
	
	\item We propose an IMP-GCN model which exploits high-order neighbors from the same subgraph, in which the user nodes share more similar interests than those in other subgraphs. It is proved to be effective on alleviating the over-smoothing problem. 
	\item We design a subgraph generation module to group users and generate subgraphs from the user-item bipartite graph by considering users features and graph structure information. 
	
	\item We conduct empirical studies on three benchmark datasets to evaluate the proposed IPM-GCN model. Results show that IPM-GCN can gain improvement by stacking more layers and learn better user/item embeddings, and thus outperforms the SOTA GCN-based recommendation models with a large margin.
\end{itemize}
\section{METHODOLOGY}
\label{sec:methods}
\subsection{Recap}
Let $\bm{A} \in \mathds{R}^{N \times M}$ be the user-item interaction matrix, where $N$ and $M$ indicate the number of users and items, respectively. An nonzero entry $a_{ui}\in \bm{A}$ indicates that user $u \in \mathcal{U}$  has interacted with  item ${i} \in \mathcal{I}$ before; otherwise, the entry is zero. A user-item bipartite graph $\mathcal{G}= (\mathcal{W}, \mathcal{E})$ can be constructed based on the interaction matrix, where the node set $\mathcal{W}$ consists of the two types of user nodes and item nodes and   $\mathcal{E}$ represents for the set of edges. For a nonzero  $a_{ui}$, there is an edge between the user $u$ and item $i$. The above information is taken as the input of GCN model to learn the user and item representations by iteratively aggregating features from neighboring nodes in the bipartite graph.

Here we take LightGCN as an example to describe the GCN-based recommendation model, because it achieves the state-of-the-art performance with a very light design. Our model is also developed based on its design.\footnote{Note that although LR-GCN was inspired by a different motivation, its final formulation is almost the same as LightGCN.} Let $\bm{e_u^{(0)}}$ denote the ID embedding of user $u$ and $\bm{e_i^{(0)}}$ denote the ID embedding of item $i$, the graph convolution operation in LightGCN is described as follows:
\begin{equation}
\begin{aligned}
\label{equation:First-order}
\bm{e_u^{(k)}} = \sum_{i \in \mathcal{N}_u}\frac{1}{\sqrt{ | \mathcal{N}_u | }\sqrt{ | \mathcal{N}_i |}}\bm{e_i^{(k-1)}}, \\
\bm{e_i^{(k)}} = \sum_{u \in \mathcal{N}_i}\frac{1}{\sqrt{ | \mathcal{N}_i | }\sqrt{ | \mathcal{N}_u |}}\bm{e_u^{(k-1)}},
\end{aligned}
\end{equation}
where $\bm{e_u^{(k)}}$ and $\bm{e_i^{(k)}}$ represent the embeddings of the user $u$ and item $i$ after $k$ layers propagation, respectively; $\mathcal{N}_u$ denotes the set of items that interact with user $u$, and $\mathcal{N}_i$ denotes the set of users that interact with item $i$; $\frac{1}{\sqrt{ | \mathcal{N}_u | }\sqrt{ | \mathcal{N}_i |}}$ is symmetric normalization terms, which can avoid the scale of embeddings increasing with graph convolution operations~\cite{kipf2017gcn}. After K layers graph convolution, the final embeddings of a user $u$ and an item $i$ are the combination of  their embeddings obtained at each layer in LightGCN:
%Existed GCN-based recommendation models always suffer oversmooting problems.
\begin{equation}
\label{final_representation}
\bm{e_u} = \sum_{k=0}^{K}\alpha_{k}\bm{e_u^{(k)}};
\bm{e_i} = \sum_{k=0}^{K}\alpha_{k}\bm{e_i^{(k)}},
\end{equation}
where $\alpha_{k} \geq 0$ is a hyper-parameter assigned to the k-th layer. It denotes the importance of this layer in constituting the final embedding. From Eq.~\ref{final_representation}, it is expected that after iteratively aggregating features from higher-order neighbors, the nodes will fail to preserve their own distinct features and their embeddings become more and more similar, leading to the over-smoothing problem. Besides, it does not distinguish the heterogeneous features of high-order nodes in the aggregation process. The noisy information  from high-order neighbors could hurt the embedding learning. For example, the embeddings of users with no common interests or even contradictory interests in the high-order neighbors are aggregated to learn a target user's embedding via the graph convolution operation. 

\begin{figure}[t]
	\centering
	\hspace{-0.0cm}
	\includegraphics[width=0.9\linewidth]{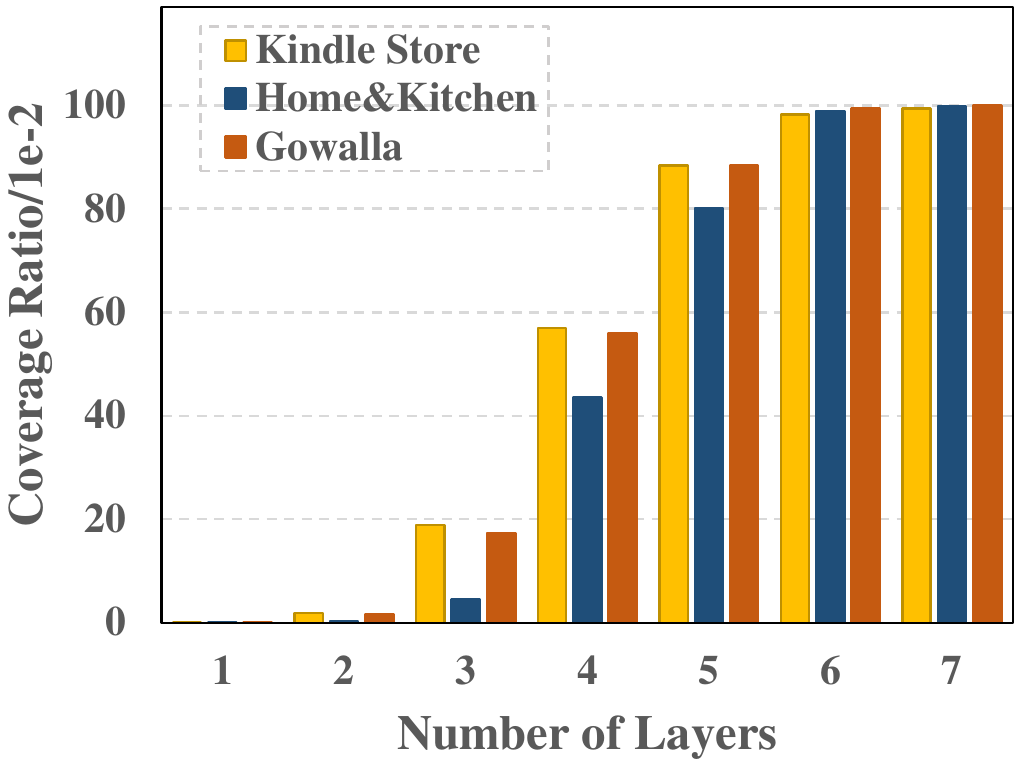}
	\vspace{-0pt}
	\caption{The average ratio of nodes involved in different layers of graph convolution on three datasets.}
	\vspace{-5pt}
	\label{fig:sec1_coverage_graph}
\end{figure}
Fig.~\ref{fig:sec1_coverage_graph} shows the average coverage ratio of \textit{the number of nodes that a target node reaches in the propagation by stacking different numbers of layers} to \textit{all the nodes in the graph}. It can be seen that after 6- or 7-layer graph convolution, a node can almost receive information from all the other nodes in embedding propagation. Therefore, by aggregating information from all the connected high-order neighbors, it is unavoidable that the node embeddings become homogeneous in the current GCN-based models after stacking more layers, especially for the densely connected ones, whose embeddings will become more and more similar. In the recommendation scenario, this means the uniqueness of users will be neglected in deep structure.

Actually, current GCN-based recommendation models achieve their peak performance at most 3 or 4 layers~\cite{He2020lightgcn, pmlr-v97-wu19e}. Besides the over-smoothing effect, we deem that a node also takes noisy or negative information in the embedding propagation process, which hurts the final performance. This is because a user's interests often span a range of items. Different users can have very different interests or even exhibit contradictory attitudes to some items. Without distinguishing those users, the embedding propagation may perform among users with very different interests to learn their embeddings in the graph convolution operation. To avoid the situation and alleviate the over-smoothing problem, it is important to group users with similar interests (and their interacted items) into subgraphs and constrain the embedding propagation to operate inside the subgraph. To achieve the goal we propose the interest-aware message-passing GCN model. 
\subsection{IMP-GCN MODEL}
\begin{figure*}[t]
	\centering
	\includegraphics[width=1.0\linewidth]{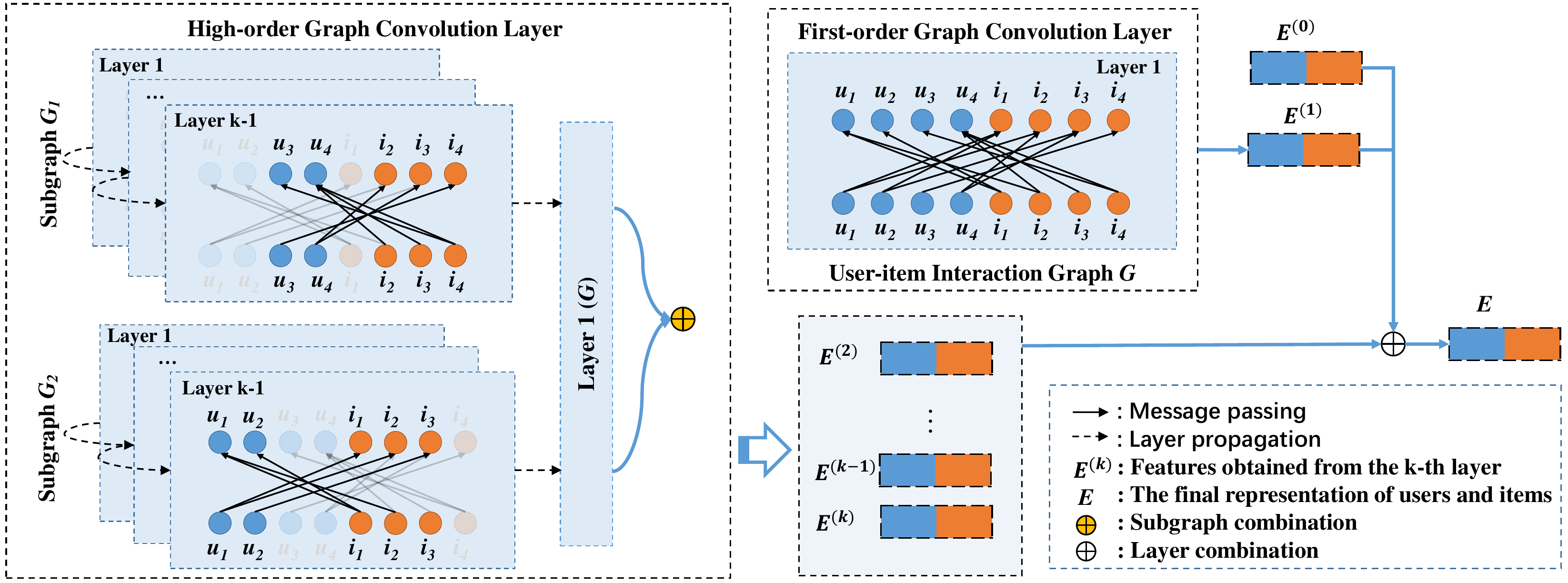}
	\caption{An overview of our IMP-GCN model with two subgraphs as illustration. In IMP-GCN, the first-order  propagation operates on whole graph, and high-order propagation operates inside the subgraphs.}
	\vspace{-5pt}
	\label{fig:sec2_message_passing}
\end{figure*}
\subsubsection{Interest-aware Message-passing Strategy}
\label{section:FGCL}
With constructing subgraphs, we would like that all the information propagated in a subgraph can contribute to the embedding learning of all the nodes in this subgraph. In other words, we aim to exclude the negative information propagation in the graph convolution operation using subgraphs. To achieve the goal, we rely on user nodes to form subgraphs in the user-item bipartite graph. The general idea is that users with more similar interests are grouped into a subgraph, and the items which directly linked to those users also belong to this subgraph. Therefore, each user  only belongs to one subgraph, and an item can be associated with multiple subgraph. Let $G_{s}$ with $s \in\{1,\cdots,N_{s} \}$ denotes a subgraph, where $N_s$ is the number of subgraphs. In the next, we introduce the graph convolution operation in our model.

Because the direct interactions between users and items  provide the most important and reliable information of user interests, in the first-order propagation, all the first-order neighbors are involved in the graph convolution operation.  Let $\bm{e_u^{(0)}}$ and $\bm{e_i^{(0)}}$ denote the ID embeddings of user $u$ and item $i$, respectively. The first-order graph convolution is:
\begin{equation}
\label{first_order_MP}
\begin{aligned}
\bm{e_{u}^{(1)}} =  \sum_{i \in \mathcal{N}_{u}}\frac{1}{\sqrt{ | \mathcal{N}_{u} | }\sqrt{ | \mathcal{N}_{i} |}}\bm{e_{i}^{(0)}}, \\
\bm{e_{i}^{(1)}} =  \sum_{u \in \mathcal{N}_{i}}\frac{1}{\sqrt{ | \mathcal{N}_{i} | }\sqrt{ | \mathcal{N}_{u} |}}\bm{e_{u}^{(0)}}, 
\end{aligned}
\end{equation}
where  $\bm{e_u^{(1)}}$  and $\bm{e_i^{(1)}}$ represent the first layer embeddings of the target user $u$ and item $i$, respectively. 

For the high-order graph convolution, to avoid introducing noisy information, a node in a subgraph can only exploit the information from its neighbor nodes in this subgraph. Because the items interacted by a user all belong to the subgraph of this user, the user can still receive information from all the linked items. However, for an item node, its direct user neighbors can be distributed in different subgraphs. To learn the embeddings of an item $i$, for each subgraph $G_s$ it belongs to, we learn an embedding for this item. Let $\bm{e_{is}^{(k)}}$ denotes the embedding of item $i$ in subgraph $s$ after $k$ layers graph convolution, the high-order propagation in IMP-GCN is defined as:
\begin{equation}
\begin{aligned}
\bm{e_{u}^{(k+1)}} =  \sum_{is \in \mathcal{N}_{u}}\frac{1}{\sqrt{ | \mathcal{N}_{u} | }\sqrt{ | \mathcal{N}_{i} |}}\bm{e_{is}^{(k)}}, \\
\bm{e_{is}^{(k+1)}} =  \sum_{u \in \mathcal{N}_{i}^s}\frac{1}{\sqrt{ | \mathcal{N}_{i} | }\sqrt{ | \mathcal{N}_{u} |}}\bm{e_{u}^{(k)}}.
\end{aligned}
\end{equation}
In this way, we guarantee that the embedding of a node learned in a subgraph only contributes to the embedding learning of other nodes in this subgraph. This can avoid the noisy information propagated from unrelated nodes.  $\bm{e_{is}^{(\cdot)}}$ can be regarded as the features learned from the users with a similar interest in the subgraph $G_s$. This make senses since users with similar interests often prefer the same feature of an item. The final representation of an item $i$ after $k$ layers graph convolution is a combination of its embeddings learned in different subgraphs, i.e., 

\begin{equation}
\label{equation:message_passing_2}
\bm{e_i^{(k)}} = \sum_{s \in \mathcal{S}}\bm{e_{is}^{(k)}},
\end{equation} 
where $S$ is the subgraph set that item $i$ belongs to. 

\subsubsection{Layer Combination and Prediction}
We combine the embeddings obtained at each layer to form the final representation of user $u$ and item $i$ as Eq.~\ref{final_representation}. Similar to LightGCN, $\alpha_{k}$ is set uniformly as $1/(K + 1)$~\cite{He2020lightgcn}. 

With the learned embeddings of users (i.e., $\bm{e}_{u}$) and items $\bm{e}_{i}$, given a user $u$ and a target item $i$, the preference of the user to the item is computed by inner product:
\begin{equation}
\hat{r}_{uv} = \bm{e_{u}}^T\bm{e_{i}}.
\end{equation}
Notice that other interaction functions can be also applied, such as Euclidean distance. Because the main focus of this work is to study the effects of distinguishing user interests in the graph convolution in the GCN-based recommendation model, we adopt the inner product as previous work~\cite{hoprec,chen2019cse,wang2019ngcf} for fair comparisons in the empirical studies.

\subsubsection{Matrix-form propagation rule.}
We implement our algorithm with the matrix form propagation rule (see~\cite{wang2019ngcf} for more details), by which we can simultaneously update the representations of all users and items in a rather efficient way. It is a commonly used approach to make graph convolution network feasible for large-scale graph~\cite{wang2019ngcf,qiu2018deepinf}. Let $\bm{E}^{(0)}$ be the representations matrix for users ID and items ID; $\bm{E^{(k)}}$ represents the representation of users and items at the $k$-th layer. Similarly, $\bm{E_s^{(k)}}$ is defined as the representation of users and items at the $k$-th layer in subgraph $G_s$. As shown in Fig.~\ref{fig:sec2_message_passing},
the first layer embedding propagation in our model can be described as follows:
\begin{equation}
\bm{E^{(1)}} = \mathcal{L}\bm{E^{(0)}}, 
\end{equation}
where $\mathcal{L}$ is the Laplacian matrix for the user-item interaction graph.

As we involve the subgraphs in high-order graph convolution layers,the embeddings propagation on subgraphs is formulated as follows:
\begin{equation}
\bm{E_s^{(k-1)}} =  \mathcal{L}_{s}\bm{E_s^{(k-2)}},
\end{equation}
where $k \geqslant 2$; $\mathcal{L}_{s}$ represent the Laplacian matrix for the subgraph $G_{s}$. And then, the $(k-1)$-th layer embeddings are propagated on the user-item graph and obtained the embeddings in the $k$-th layer:
\begin{equation}
\bm{E_s^{(k)}} = \mathcal{L}\bm{{E_s}^{(k-1)}}. 
\end{equation}
We aggregate all the $k$-th layer embeddings involved different subgraphs to formulate the final $k$-th layer embeddings:
\begin{equation}
\bm{E^{(k)}} = \sum_{s \in G_s}\bm{E_{s}^{(k)}}.
\end{equation}
Lastly, we combine all the layers' embeddings and get the final representations of users and items, this formulation keeps consistent with it in LightGCN~\cite{He2020lightgcn}:
\begin{equation}
\bm{E} = \alpha_{0}\bm{E^{(0)}} + \alpha_{1}\bm{E^{(1)}} + \cdots + \alpha_{K}\bm{E^{(K)}}
\end{equation}

\subsubsection{Optimization}
In this work, we target at the top-$n$ recommendation, which aims to recommend a set of $n$ top-ranked items matching the target user's preference. Compared to rating prediction, this is a more practical task in real commercial systems~\cite{rendle2009bpr}. Similar to other rank-oriented recommendation works~\cite{hoprec,wang2019ngcf}, we adopt the pairwise learning method for optimization. To perform the pairwise learning, it needs to constructs a triplet of $\{u, i^+, i^-\}$, with \textit{an observed interaction between $u$ and $i^+$} and \textit{an unobserved interaction between $u$ and $i^-$}.  This method assumes that a positive item (i.e., $i^+$) should rank higher than an negative item (i.e., $i^-$). The objective function is formulated as:
\begin{equation}
\mathop{\arg\min} \sum_{(\mathbf{u}, \mathbf{i}^+,\mathbf{i}^-)\in{\mathcal{O}}} -\ln\phi(\hat{r}_{ui^+} - \hat{r}_{ui^-}) + \lambda\left\|\Theta\right\|^2_2
\end{equation}
where $\mathcal{O}=\{(u, i^+, i^-)|(u,i^+)\in\mathcal{R^+}, (u,i^-) \in\mathcal{R^-}\}$ denotes the training set;  $\mathcal{R^+}$ indicates the observed interactions between user $u$ and $i^+$ in the training dataset, and $\mathcal{R^-}$ is the sampled unobserved interaction set. $\lambda$ and $\Theta$ represent the regularization weight and the parameters of the model, respectively. The $L_2$ regularization is used to prevent overfitting.

The mini-batch Adam~\cite{kingma2014adam} is adopted to optimize the prediction model and update the model parameters. Specifically, for a batch of randomly sampled triples $(u, v^+, v^-)\in \mathcal(O)$, the representation of those users and items are first learned by the propagation rules and then the model parameters are updated by using the gradients of the loss function. 

\subsection{Subgraph Generation Module}
\label{sec:SGM}
In this section, we introduce our proposed subgraph generation module which is designed to construct the subgraphs $G_{s}$ with $s \in\{1,\cdots,N_{s} \}$ from a given input graph $\mathcal{G}$. Remind that the subgraphs are used to group users with common interests in our model. We formulate the user grouping as a classification task~\cite{hu2020temporal}, i.e., each user is classified to a group. Specifically, each user is represented by a feature vector, which is a fusion of the graph structure and the ID embedding:
\begin{equation}
\label{feature_fusion}
\bm{F_{u}} = \sigma (\bm{W_{1}}(\bm{e_u^{(0)}} + \bm{e_u^{(1)}}) + \bm{b_{1}}),
\end{equation}
where $\bm{F_{u}}$ is the obtained user feature via feature fusion. $\bm{e_u^{(0)}}$ is the embedding of user ID and $\bm{e_u^{(1)}}$ is the feature obtained by aggregating local neighbor in the graph (i.e., the user embedding after the first layer propagation.). $\bm{W_{1}} \in R^{d \times d}$ and $\bm{b_1} \in R^{1 \times d}$ are respectively the weight matrix and bias vector of the fusion method. $\sigma$ is the activation function. LeakyReLU~\cite{Andrew2013leaky} is adopted, because it can encode both positive and small negative signals. To classify the users into different subgraphs, we cast the obtained user feature to a prediction vector with a 2-layer neural networks:
\begin{gather} \label{eq:group}
\begin{split}
\bm{U_{h}} = \sigma(\bm{W_{2}F_{u}} + \bm{b_{2}}),\\
\bm{U_{o}} = \bm{W_{3}U_{h}} + \bm{b_3},
\end{split}
\end{gather}
where $\bm{U_{o}}$ is the prediction vector. The position of maximum value in $U_o$ represents which group/subgraph the user belongs to. $\bm{W_{2}} \in R^{d \times d}$, $\bm{W_{3}} \in R^{d \times N_{s}}$ and $\bm{b_2} \in R^{1 \times d}$, $\bm{b_3} \in R^{1 \times N_{s}}$ are respectively the weight matrices and bias vectors of the two layers. The dimension of the prediction vector dimensions is the same as the number of subgraphs, which is a pre-selected hyper-parameter. Note that it is an unsupervised method to classify users into different groups and thus does not need ground-truth label. For users with similar embeddings, Eq.~\ref{eq:group} will generate similar prediction vector, namely, they will be classified into the same group. The subgraph generation aims to construct a matrix, which represents the user-item adjacency relation in a subgraph based on the user grouping results and the Laplacian matrix of the original user-item graph. For the matrix of each subgraph, according to the obtained user group information, we filter out the user-item adjacency relations in the Laplacian matrix of the original user-item graph if the corresponding users are not in the user group.
\section{Experiments}
\label{sec:experiments}
\subsection{Experimental Setup}
\subsubsection{Data Description}
\label{sec:data_describe}
\begin{table}[t]
	\centering
	\caption{ Basic statistics of the experimental datasets.}
	%\vspace{-0.2cm}
	\label{tab:data}
	%\resizebox{\textwidth}{!}{
	\begin{tabular}{ccccc}
		\toprule
		Dataset&\#user&\#item&\#interactions&sparsity \\
		\midrule
		Kindle Store & 68,223& 61,934 & 982,618 &99.98\% \\
		Home\&Kitchen & 66,519 & 28,237 & 551,681 & 99.97\% \\
		Gowalla & 29,858 & 40,981 & 1,027,370 & 99.92\% \\
		\bottomrule
	\end{tabular}
	\vspace{-5pt}
\end{table}
To evaluate the effectiveness of IMP-GCN, we conducted experiments on three benchmark datasets: Amazon-Kindle Store, Amazon-Home$\&$Kitchen and Gowalla. The first two datasets are from the public Amazon review dataset\footnote{http://jmcauley.ucsd.edu/data/amazon.}, which has been widely used for recommendation evaluation in previous studies. The third dataset is a check-in dataset collected from Gowalla, where users share their locations by checking-in.
We followed the general setting in recommendation to filter users and items with few interactions. For all the datasets, we used the 10-core settings, i.e., retaining users and items with at least 10 interactions. The statistics of three datasets are shown in Table~\ref{tab:data}. As we can see, the datasets are of different sizes and sparsity levels, which are useful for analyzing the performance of our method and the competitors in different situations. 

For each datasets, we randomly split it into training, validation, and testing set with the ratio 80:10:10 for each user. The observed user-item interactions were treated as positive instances. For the methods which adopt the pairwise learning strategy, we randomly sample a negative instance, that the user did not consume before, to pair with each positive instance.
\begin{figure*}[t]
	\centering
	\includegraphics[width=1.0\linewidth]{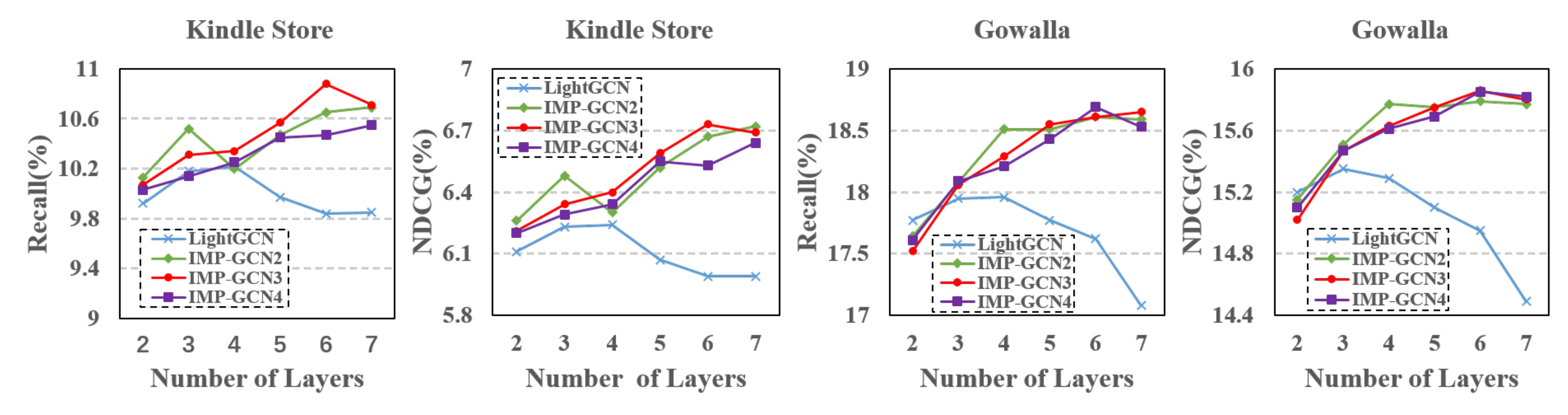}
	\vspace{-0pt}
	\caption{Results Comparison between IMP-GCN  and LightGCN at different layers on Kindle Store and Gowalla. IMP-GCN$_{2}$, IMP-GCN$_{3}$, and IMP-GCN$_{4}$ represent IMP-GCN with 2, 3, and 4 subgraphs, respectively.}
	\vspace{-0pt}
	\label{fig:sec4_layers_groups}
\end{figure*}
\subsubsection{Evaluation Metrics}
For each user in the test set, we treat all the items that the user did not interact with as negative items. Two widely used evaluation metrics for top-$n$ recommendation are adopted in our evaluation: Recall and Normalized Discounted Cumulative Gain~\cite{he2015trirank}.
For each metric, the performance is computed based on the top 20 results. Notice that the reported results are the average values across all the testing users.
\subsubsection{Experimental Settings}
We implemented our model with Tensorflow~\footnote{https://www.tensorflow.org.} and carefully tuned the key parameters. The embedding size is fixed to 64 for all models and the embedding parameters are initialized with the Xavier method~\cite{HuY2019, Xavier2010xavier}. We optimized our method with Adam~\cite{kingma2014adam} and used the default learning rate of 0.001 and default mini-batch size of 1024 (on gowalla, we increased the mini-batch size to 2048 for speed). The $L_{2}$ regularization coefficient $\lambda$ is searched in the range of $\{1e^{-6},1e^{-5},\cdots,1e^{-2}\}$. The early stopping and validation strategies are kept the same as those in LightGCN.
\subsection{Study of IMP-GCN}
In this section, we first evaluated the performance of our IPM-GCN model when stacking different layers in graph convolution. This is to examine whether our interest-aware message-passing strategy can alleviate the over-smoothing problem. In the next, we study the effects of the subgraph numbers on the performance of our model. 
\subsubsection{Effect of Layer Numbers}
To investigate the effectiveness of IMP-GCN in deeper structure, we increased the model depth and performed detailed comparison with LightGCN. Since the adopted message-passing strategy is the same as LightGCN in the first-order convolution layer, we increased the layer number from 2 to 7. The experimental results are shown in Fig.~\ref{fig:sec4_layers_groups}, in which IMP-GCN$_{2}$, IMP-GCN$_{3}$ and IMP-GCN$_{4}$ indicate the model with 2, 3, and 4 subgraphs, respectively. We omitted the results on $Home\&Kitchen$ for space limitation, because they show exactly the same trend. From the results, we had some interesting observations.

Firstly, the proposed IMP-GCN outperforms LightGCN consistently when stacking more than 2 or 3 layers over both datasets. This indicates that our model can learn better embeddings by the interest-aware message-passing strategy. Secondly, the peak performance of LightGCN is obtained when stacking 3 or 4 layers, and increasing more layers will cause dramatic performance degradation, indicating it suffers from the over-smoothing problem in a deep structure. In contrast, IMP-GCN continues to achieve better performance with deeper structure (notice that when stacking more than 7 layers, a node already aggregates information from almost all the nodes, see Fig. ~\ref{fig:sec1_coverage_graph}. The results demonstrate the capability of our model on alleviating the over-smoothing problem. Moreover, it also 1) justifies our claim that exploiting information from all nodes indiscriminately causes the over-smoothing in GCN-based recommendation model, and 2) validates the effectiveness of our subgraph generation algorithm on classifying users with common interests.
\begin{figure}[t]
	\centering
	\subfloat[Recall on Kindle Store]{\includegraphics[width=1.6in]{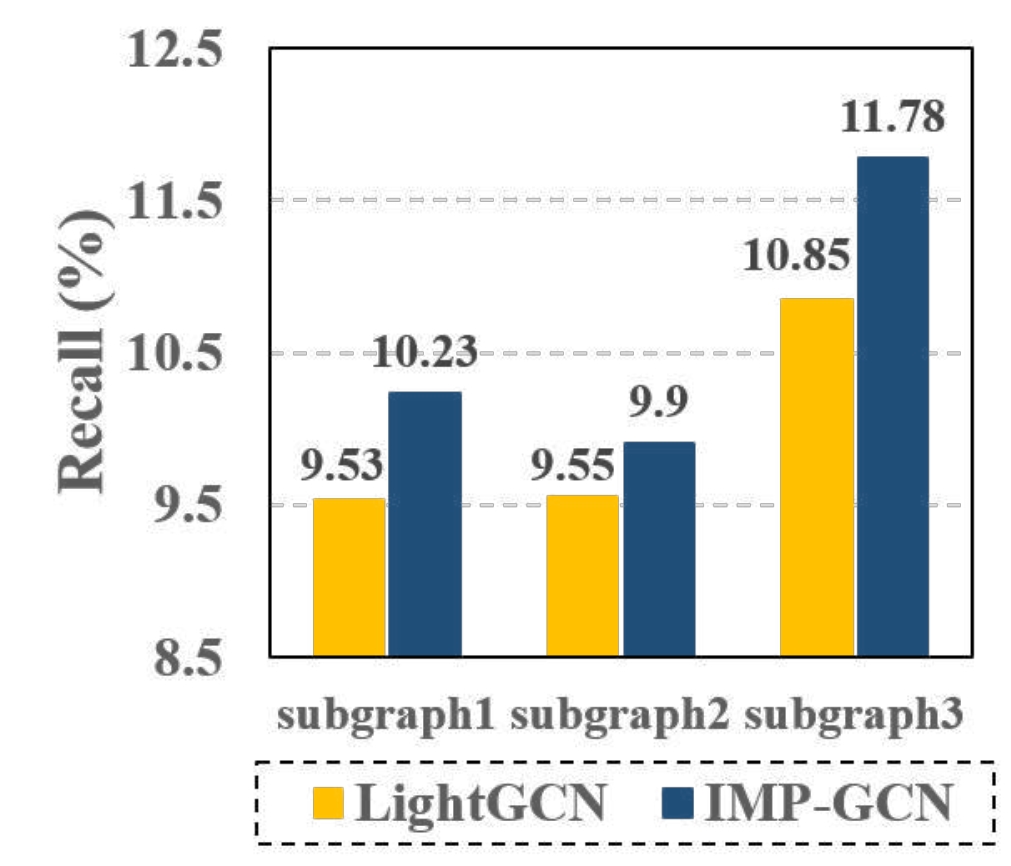}}
	\subfloat[Coverage Ratio on Kindle Store]{\includegraphics[width=1.6in]{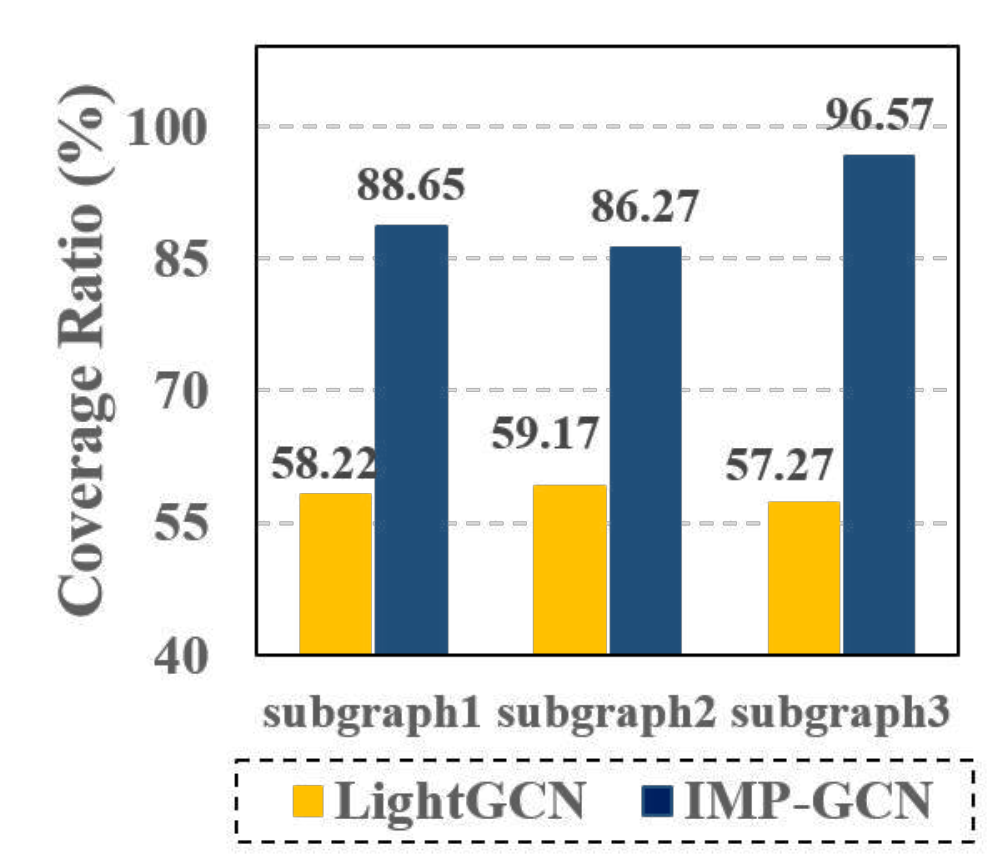}}
	\vspace{-0pt}
	\caption{Statistics of Recall and Coverage Ratio on Kindle Store in three subgraphs.}
	\vspace{-10pt}
	\label{fig:sec4_cv_recall_groups}
\end{figure}
\subsubsection{Effect of Subgraph}
The performance of IPM-GCN with different numbers (i.e., $\{2, 3, 4\}$) of subgraphs can also be observed in Fig.~\ref{fig:sec4_layers_groups}. From the results, we can see that the (1) IMP-GCN$_2$ with 2 subgraphs can obtain the best results when stacking no more than 3 layers. This is because a node in the subgraphs of IMP-GCN$_2$ can reach more nodes in short distance than the on in IMP-GCN$_3$ or IMP-GCN$_4$in the embedding propagation operation.
(2) When stacking more than 3 layers, IMP-GCN$_3$ performs the best. After 3 layers graph convolution, the number of involved nodes increasing sharply in embedding propagation (see the examples in Fig.~\ref{fig:sec1_coverage_graph}). On average, each node in IMP-GCN$_2$ should reach more nodes that the one in IMP-GCN$_3$ and IMP-GCN$_4$, however, the performance improvement of IMP-GCN$_2$ is smaller or even negative (on th Kindle Stores)  than that of IMP-GCN$_3$ and IMP-GCN$_4$. This indicates that  there is still noisy information in embedding propagation by discriminating user interests in a coarse-level (i.e., 2 subgraphs), negatively impacting the performance. Note that IMP-GCN$_3$ can still benefit from high-order neighbors.
(3) With more subgraphs, on the one hand, IMP-GCN$_4$ can distinguish users with similar interests in a finer level and thus can better distill information from high-order neighbors; on the other hand, it also cuts more connections to other nodes, especially the ones in short distance which provide  more valuable information in embedding learning. As a result, when stacking more layers,  its performance is only comparable to that of IMP-GCN$_2$. Therefore, there is a trade-off on selecting the number of subgraphs. 
We further studied the effects of subgraphs by analyzing the average coverage ratio of each node and the corresponding performance based on the LightGCN and our IPM-GCN model.  Due to the space limitation, we only provide the results on Kindle Store and omit the performance $w.r.t$ ndcg which has the similar trend as recall. In this experiment, we used the LightGCN with 4 layers and IPM-GCN with 3 subgraphs\footnote{Number of users in the three groups $G_1$, $G_2$, $G_3$ are $3,971$, $3,584$, $6,801$, respectively.} and 6 layers, which are their optimal setting on Kindle Store. The average recall and average cover ratio of each user in a subgraph  based on LightGCN and IPM-GCN are shown in Fig.~\ref{fig:sec4_cv_recall_groups}(a) and Fig.~\ref{fig:sec4_cv_recall_groups}(b), respectively. Notably, by grouping users with similar interest in subgraphs to make information only propagate inside subgraphs, IPM-GCN can benefit from more layers of graph convolution and distill positive information from high-order neighors. In contrast, LightGCN is limited by the negative information from high-order neighbors and can only gain improvements over 4 layers. Comparing the performance of different subgraphs, we can see that with a higher coverage ratio, the performance of IPM-GCN increases clearly. 

Another interesting finding is that, by stacking 6 layers, a user node in a subgraph almost connects to all the other nodes in the whole graph. This indicates that the users in a subgraph almost interact all the items in the graph (otherwise, the coverage ratio cannot be that high). More importantly, IPM-GCN can still achieve improvement with such high coverage without over-smoothing. This indicates that the embeddings of items learned in a graph contributes to the embedding learning of  users in this graph, and the distilled information in a subgraph during graph convolution is useful for the embedding learning for all the nodes in this subgraph. It demonstrates the effectiveness of our interest-aware message-passing strategy and the subgraph generation algorithm.
\subsection{Comparison with SOTA Methods}
\subsubsection{Baselines}
To demonstrate the effectiveness, we compared our proposed method with several recently proposed competitive methods, including
\begin{itemize}[leftmargin=*]
	\item \textbf{NeuMF~\cite{he2017neural}}: It is a state-of-the-art neural collaborative filtering method. This method uses multiple hidden layers above the element-wise and concatenation of user and item embeddings to capture their non-linear feature interactions. 
	
	\item \textbf{HOP-Rec~\cite{hoprec}:} This method exploits the high-order user-item interactions by random walks to enrich the original training data. In experiments, we used the codes released by the authors~\footnote{https://github.com/cnclabs/smore.}.
	
	\item \textbf{CSE~\cite{chen2019cse}:} This recently proposed graph-based model also exploits the high-order proximity in the user-item bipartite graph. Different from HOP-Rec, this method explores the user-user and item-item relations by random walks to improve the performance. We used the codes released by the authors ( the same link as HOP-Rec). 
	
	\item \textbf{GCMC~\cite{berg2019gcmc}:} This method applies the GCN techniques on user-item bipartite graph and employs one convolutional layer to exploit the direct connections between users and items. 
	
	\item \textbf{NGCF~\cite{wang2019ngcf}}: This method explicitly encodes the collaborative signal in the form of high-order connectivities by performing embedding propagation in the user-item bipartite graph.
	
	\item \textbf{LightGCN~\cite{He2020lightgcn}}: It is an simplified version of NGCF by removing the feature transformation and nonlinear activation module. It makes GCN-based methods more concise and appropriate for recommendation and achieves the state-of-the-art performance.
\end{itemize}
For fair comparisons, all the methods are optimized by the same pairwise learning strategy. We put great efforts to tune these methods based on the validation dataset and reported their best performance.
\subsubsection{Overall Comparison}
\begin{table}[t]
	\vspace{0pt}
	\caption{Performance of our model and the competitors over three datasets. Noticed that the values are reported by percentage with '\%' omitted. } 
	\centering
	\resizebox{0.5\textwidth}{!}{
		\begin{tabular}{l|cc|cc|cc} \hline
			Datasets	& \multicolumn{2}{c|}{Kindle Store} & \multicolumn{2}{c|}{Home\&Kitchen} & \multicolumn{2}{c}{Gowalla} \\ \cline{2-3}  \cline{4-5} \cline{6-7}
			Metrics	& Recall	&	NDCG	& Recall	& NDCG	& Recall    &	NDCG	\\ \hline \hline
			NeuMF	&	4.96	&	2.06    & 1.34      & 0.62	&	12.96	&	11.21  \\ \hline
			CSE	    &	7.65	&	4.54    & 1.93      & 0.91	&	13.85	&	11.51  \\
			HOP-Rec	&	7.96	&   4.58    & 1.98      & 0.94	&	14.11	&	12.70  	\\ \hline
			GCMC	&	7.93	&   4.55    & 1.42      & 0.64	&	14.03	&	11.68 \\
			NGCF	&	8.25	&   5.09	& 2.14      & 0.96  &	15.62	&	13.35  	\\ 
			LightGCN &	\textbf{10.22}	& \textbf{6.24} 	& 	\textbf{3.03} 	& \textbf{1.39}  & \textbf{17.96}   & \textbf{15.29}	  	\\ \hline \hline
			IMP-GCN	 &	\textbf{10.88*}	& \textbf{6.73*}	&	\textbf{3.22*}	& \textbf{1.49*} & \textbf{18.69*}	& \textbf{15.85*}		\\ \hline
			Improv.  & 6.46\% & 7.85\% & 6.27\% & 7.19\% & 4.07\% & 3.66\% 
			\\ \hline 
	\end{tabular}}
	\begin{tablenotes}
		\footnotesize
		\item The symbol * denotes that the improvement is significant with $p-value < 0.05$ based on a two-tailed paired t-test.
	\end{tablenotes}
	\label{tab:results}
	\vspace{-2pt}
\end{table}
Table~\ref{tab:results} shows the performance comparison results. The best and second best results were highlighted in bold. From the results, we had following observations.

The performance of NeuMF is relatively poor as it not explicitly leverages the high-order connectivities between users and items, resulting in suboptimal performance. For the graph-based methods, CSE makes use of the implicit associates of user-user and item-item similarities via high-order neighborhood proximity by performing random walks on the user-item interaction graph. GCMC obtains better performance over CSE, demonstrating the advantages of GCN-based approaches, which can exploit graph structure information. However, it does not perform well on $Home\&Kitchen$ because the useful information in neighbors cannot be efficiently aggregated. Hop-Rec outperforms the above methods on the three datasets, because it samples user-item interactions from high-order neighbors to enrich the training data. NGCF achieves consistent much better performance over the above baselines. This is because it adopts the GCN techniques to explicitly and directly exploit the high-order connectivities in the embedding. In contrast, the GCMC method only utilizes the first-order neighbors for representation learning; HOP-Rec and CSE leverage the high-order neighbors to enrich the training data rather than using them in embedding function for direct representation learning. This demonstrates the powerful representation learning capability of GCN and the importance of utilizing high-order information directly in representation learning. Similar to the results reported in~\cite{He2020lightgcn}, LightGCN achieves substantially improvement over NGCF by simplifying it with the removal of two common designs in GCN.

IMP-GCN outperforms all the baselines consistently over all the datasets. In particular, compared to the strongest baseline in terms of NDCG@20, IMP-GCN can reach a relative improvement over LightGCN by 7.85\%, 7.19\%, 3.66\% on $Kindle Store$, $Home\&Kitchen$ and $Gowalla$, respectively. The great improvement over LightGCN demonstrates the importance of distinguishing nodes in high-order neighbors in the graph convolution operation, as well as the effectiveness of our proposed interest-aware message-passing strategy. 
\begin{table}[]
	\caption{Performance of our model and its variants over three datasets. Noticed that the values are reported by percentage with '\%' omitted. } 
	\centering
	\resizebox{0.5\textwidth}{!}{
		\begin{tabular}{l|cc|cc|cc} \hline
			Datasets	& \multicolumn{2}{c|}{Kindle Store} & \multicolumn{2}{c|}{Home\&Kitchen} & \multicolumn{2}{c}{Gowalla} \\ \cline{2-3}  \cline{4-5} \cline{6-7}
			Metrics	& Recall	&	NDCG	& Recall	& NDCG	& Recall    &	NDCG	\\ \hline \hline
			IMP-GCN$_s$	&	10.57	&   6.63	& 3.14      & 1.43  &	18.61	&	15.61  	\\ 
			IMP-GCN$_f$ &	10.19	& 6.40 	& 	2.97 	& 1.31  & 17.84   & 15.11	  	\\  
			IMP-GCN	 &	\textbf{10.88}	& \textbf{6.73}	&	\textbf{3.22}	& \textbf{1.49} & \textbf{18.69}	& \textbf{15.85}	\\ \hline
	\end{tabular}}
	\label{tab:ablation_results}
	\vspace{-10pt}
\end{table}
\subsection{Ablation Study}
In this section, we examined the contribution of different components in our model to the final performance by comparing IMP-GCN with the following two variants:
\begin{itemize}[leftmargin=*]
	\item \textbf{IMP-GCN$_{s}$}: This variant removes the graph structure information from the subgraph generation module (i.e., removing $\bm{e_u^{(1)}}$ in Eq.~\ref{feature_fusion}).
	\item \textbf{IMP-GCN$_{f}$}: In this variant, the first-order propagation is also performed inside each subgraph (i.e., The equation for $\bm{e_i^{(1)}}$ in Eq.~\ref{first_order_MP} is replaced with $\sum_{s \in \mathcal{S}}\bm{e_{is}^{(1)}}$). 
\end{itemize}

The results of two variants and IPM-GCN were reported in Table~\ref{tab:ablation_results}, in which the best results are highlighted in bold. IMP-GCN outperforms IMP-GCN$_s$ over all the datasets, which indicates the effectiveness of employing graph structure information in subgraph generation module. It is expected that IMP-GCN$_s$ obtains much better performance over IMP-GCN$_f$, because the first-order neighbors (i.e., the interaction between users and items) contributes the direct  information for user and embedding in the collaborative filtering process. The results also demonstrate the reasonable design of our IPM-GCN model.
\section{related work}
As one of the most important information retrieval techniques, recommendation has made tremendous progress in past decades. 
Among various recommendation approaches, the model-based
collaborative filtering (CF)~\cite{Koren2009MF,rendle2009bpr,hsieh2017cml,wang2019kdd,he2017neural,cheng2018aspect,wang2019ngcf,cheng2019mmalfm,He2020lightgcn,Liu2019MAML} achieves a great success and becomes the mainstream recommendation technique. CF learns user and item embeddings by reconstructing the user-item interaction matrix. Earlier research efforts mainly focus on the shallow models, such as BPR~\cite{rendle2009bpr}, CML~\cite{hsieh2017cml}, matrix factorization (MF)~\cite{Koren2009MF}. Their success motivates the development of various variants via leveraging additional information (e.g., review~\cite{mcauley2013hidden}, image~\cite{he2016VBPR}, knowledge graph~\cite{wang2019tois,wang2018ripplenet,wang2019kdd}) to deal with different tasks (e.g., context-aware~\cite{Liu2015CACF}, session-based~\cite{Liu2020KGIE}). With the rise of deep learning, it has also been widely applied in recommendation and exhibits great potential by either enhancing the user/item embedding learning or introducing non-linearity into the interaction function, promoting another peak development of recommendation technique. Many DL-based recommendation models have been proposed, such as NeuMF~\cite{he2017neural}, Wide\&Deep~\cite{cheng2016wide}, and achieved better performance over traditional models.

Another research line is graph-based recommendation, which can explicitly exploit high-order proximity between users and items.  Early approaches infer indirect preference by random walks in the graph to provide recommendation ~\cite{ItemRank2007Gori,BW2015Christoffel,fouss2007random}. The recently proposed approaches exploit the user-item bipartite graph to enrich the user-item interactions~\cite{hoprec,yu2018walkranker} and explore other types of collaborative relations, such as user-user and item-item similar ties~\cite{chen2019cse,yu2018walkranker}. For example, HOP-Rec~\cite{hoprec} uses random sample positive user-item interactions to enrich the training data by using random walks. WalkRanker~\cite{yu2018walkranker} and CSE~\cite{chen2019cse} performs random walks to explore the high-order proximity in user-user and item-item relations. As those methods rely on random walks to sample new interactions for model training, their performance heavily depends on the quality of generated interactions by random walks. As a result, these methods need carefully selection and tuning effects.

In recent years, Graph Convolution Networks (GCNs) have attracted increasing attention in recommendation due to the powerful capability on representation learning from non-Euclidean structure~\cite{berg2019gcmc,wang2019ngcf,ying2018pinsage,fan2019social,wei2019mm,wang2019kdd,He2020lightgcn,Zhang2020SMOGCF,wei2019MMGCN,wei2020GRCN,liu2020A2GCN,MEIRec2019Fan}. And then, many GCN-based recommendation models have been developed. For example, GC-MC~\cite{berg2019gcmc} employs one convolution layer to exploit the direct connections between users and items; PinSage~\cite{ying2018pinsage} combines random walks with multiple graph convolution layers on the item-item graph for Pinterest image recommendation; MEIRec~\cite{MEIRec2019Fan} utilizes metapath-guided neighbors to exploit rich structure information for intent recommendation; NGCF~\cite{wang2019ngcf} exploits high-order proximity by propagating embeddings on the user-item interaction graph; instead of implicitly capturing the high-order connectivity through the propagation embedding, SMOG-CF~\cite{Zhang2020SMOGCF} is proposed to directly capture the high-order connectivity between neighboring nodes at any order. Multi-GCCF~\cite{Sun2019MGCCF} explicitly incorporates the user-user and item-item graphs, which is built upon the user-item bipartite graph, in the embedding learning process. Inspired by the study of simplifying GCN~\cite{pmlr-v97-wu19e}, researchers also introspect the complex design in GCN-based recommendation models. He at al.~\cite{He2020lightgcn} pointed out that the two common designs feature transformation and nonlinear activation have no positive effects on the final performance, and proposed LightGCN which substantially improves the performance over NDCG. Meanwhile, Chen et al.~\cite{ChenWHZW20} also proposed to remove the nonlinearity in the network and introduced a residual network to alleviate the over-smoothing problem in existing GCN-based recommendation models. In this paper, we move a step further on this research line. We claim that the indiscriminatively exploiting the high-order neighboring nodes is also an important reason for the over-smoothing problem for GCN-based recommendation model. A typical example is that two users with contradictory interests can be also connected via a $k$-order path in the user-item interaction graph. To tackle the problem, we propose an interest-aware message-passing strategy to make the embedding propagation only happened inside a subgraph with similar interests.
\section{Conclusion}
In this work, we argued that exploiting high-order node indiscriminately would introduce negative information into the embedding propagation in the GCN-based recommendation models, causing the performance degradation when stacking more layers. We presented a IMP-GCN model which learns user and item embeddings by performing high-order graph convolution inside subgraphs. The subgraphs are formed by a designed subgraph generation algorithm that groups users with similar interests and their interacted items into the same graph. In IMP-GCN, the embedding of a node learned in a subgraph only contributes to the embedding learning of other nodes in this subgraph. In this way, IMP-GCN can effectively avoid taking the noisy information into the embedding learning. Experiments on large-scale real-world datasets demonstrate that IMP-GCN can gain improvements by stacking more layers to exploit information from higher-order neighbors, and achieve the state-of-the-art performance.  The advantages of IMP-GCN indicate the importance of distinguishing high-order neighbors on tackling the over-smoothing problem in GCN models. We believe the insights in this study can shed light on the further development of graph-based recommendation models. 
\section{Acknowledgments}
This work is supported by the National Natural Science Foundation of China, No.:61902223, No.:U1936203; the Innovation Teams in Colleges and Universities in Jinan, No.:2018GXRC014; the Shandong Provincial Natural Science Foundation, No.:ZR2019JQ23;  Young creative team in universities of Shandong Province, No.:2020KJN012.
%\begin{acks}
%To Robert, for the bagels and explaining CMYK and color spaces.
%\end{acks}
\bibliographystyle{ACM-Reference-Format}
\bibliography{main}


\begin{thebibliography}{47}

%%% the Web address.

\ifx \showCODEN    \undefined \def \showCODEN     #1{\unskip}     \fi
\ifx \showDOI      \undefined \def \showDOI       #1{#1}\fi
\ifx \showISBNx    \undefined \def \showISBNx     #1{\unskip}     \fi
\ifx \showISBNxiii \undefined \def \showISBNxiii  #1{\unskip}     \fi
\ifx \showISSN     \undefined \def \showISSN      #1{\unskip}     \fi
\ifx \showLCCN     \undefined \def \showLCCN      #1{\unskip}     \fi
\ifx \shownote     \undefined \def \shownote      #1{#1}          \fi
\ifx \showarticletitle \undefined \def \showarticletitle #1{#1}   \fi
\ifx \showURL      \undefined \def \showURL       {\relax}        \fi
% invisible to TeX
\providecommand\bibfield[2]{#2}
\providecommand\bibinfo[2]{#2}
\providecommand\natexlab[1]{#1}
\providecommand\showeprint[2][]{arXiv:#2}

\bibitem[\protect\citeauthoryear{Bell and Koren}{Bell and Koren}{2007}]%
        {netflix}
\bibfield{author}{\bibinfo{person}{Robert~M. Bell} {and}
  \bibinfo{person}{Yehuda Koren}.} \bibinfo{year}{2007}\natexlab{}.
\newblock \showarticletitle{Lessons from the Netflix prize challenge}. In
  \bibinfo{booktitle}{\emph{{SIGKDD} Explorations}}. \bibinfo{publisher}{ACM},
  \bibinfo{pages}{75--79}.
\newblock


\bibitem[\protect\citeauthoryear{Chen, Wang, Tsai, and Yang}{Chen
  et~al\mbox{.}}{2019}]%
        {chen2019cse}
\bibfield{author}{\bibinfo{person}{Chih{-}Ming Chen},
  \bibinfo{person}{Chuan{-}Ju Wang}, \bibinfo{person}{Ming{-}Feng Tsai}, {and}
  \bibinfo{person}{Yi{-}Hsuan Yang}.} \bibinfo{year}{2019}\natexlab{}.
\newblock \showarticletitle{Collaborative Similarity Embedding for Recommender
  Systems}. In \bibinfo{booktitle}{\emph{WWW}}. \bibinfo{publisher}{ACM},
  \bibinfo{pages}{2637--2643}.
\newblock


\bibitem[\protect\citeauthoryear{Chen, Wu, Hong, Zhang, and Wang}{Chen
  et~al\mbox{.}}{2020}]%
        {ChenWHZW20}
\bibfield{author}{\bibinfo{person}{Lei Chen}, \bibinfo{person}{Le Wu},
  \bibinfo{person}{Richang Hong}, \bibinfo{person}{Kun Zhang}, {and}
  \bibinfo{person}{Meng Wang}.} \bibinfo{year}{2020}\natexlab{}.
\newblock \showarticletitle{Revisiting Graph Based Collaborative Filtering: {A}
  Linear Residual Graph Convolutional Network Approach}. In
  \bibinfo{booktitle}{\emph{The Thirty-Fourth {AAAI} Conference on Artificial
  Intelligence}}. \bibinfo{publisher}{{AAAI} Press}, \bibinfo{pages}{27--34}.
\newblock


\bibitem[\protect\citeauthoryear{Cheng, Koc, Harmsen, Shaked, Chandra, Aradhye,
  Anderson, Corrado, Chai, Ispir, et~al\mbox{.}}{Cheng et~al\mbox{.}}{2016}]%
        {cheng2016wide}
\bibfield{author}{\bibinfo{person}{Heng-Tze Cheng}, \bibinfo{person}{Levent
  Koc}, \bibinfo{person}{Jeremiah Harmsen}, \bibinfo{person}{Tal Shaked},
  \bibinfo{person}{Tushar Chandra}, \bibinfo{person}{Hrishi Aradhye},
  \bibinfo{person}{Glen Anderson}, \bibinfo{person}{Greg Corrado},
  \bibinfo{person}{Wei Chai}, \bibinfo{person}{Mustafa Ispir}, {et~al\mbox{.}}}
  \bibinfo{year}{2016}\natexlab{}.
\newblock \showarticletitle{Wide \& deep learning for recommender systems}. In
  \bibinfo{booktitle}{\emph{Proceedings of the 1st Workshop on Deep Learning
  for Recommender Systems}}. \bibinfo{publisher}{ACM}, \bibinfo{pages}{7--10}.
\newblock


\bibitem[\protect\citeauthoryear{Cheng, Chang, Zhu, Kanjirathinkal, and
  Kankanhalli}{Cheng et~al\mbox{.}}{2019}]%
        {cheng2019mmalfm}
\bibfield{author}{\bibinfo{person}{Zhiyong Cheng}, \bibinfo{person}{Xiaojun
  Chang}, \bibinfo{person}{Lei Zhu}, \bibinfo{person}{Rose~C Kanjirathinkal},
  {and} \bibinfo{person}{Mohan Kankanhalli}.} \bibinfo{year}{2019}\natexlab{}.
\newblock \showarticletitle{MMALFM: Explainable recommendation by leveraging
  reviews and images}.
\newblock \bibinfo{journal}{\emph{TOIS}} \bibinfo{volume}{37},
  \bibinfo{number}{2} (\bibinfo{year}{2019}), \bibinfo{pages}{16}.
\newblock


\bibitem[\protect\citeauthoryear{Cheng, Ding, Zhu, and Mohan}{Cheng
  et~al\mbox{.}}{2018}]%
        {cheng2018aspect}
\bibfield{author}{\bibinfo{person}{Zhiyong Cheng}, \bibinfo{person}{Ying Ding},
  \bibinfo{person}{Lei Zhu}, {and} \bibinfo{person}{Kankanhalli Mohan}.}
  \bibinfo{year}{2018}\natexlab{}.
\newblock \showarticletitle{Aspect-aware latent factor model: Rating prediction
  with ratings and reviews}. In \bibinfo{booktitle}{\emph{Proceedings of the
  27th International Conference on World Wide Web}}.
  \bibinfo{publisher}{IW3C2}, \bibinfo{pages}{639--648}.
\newblock


\bibitem[\protect\citeauthoryear{Christoffel, Paudel, Newell, and
  Bernstein}{Christoffel et~al\mbox{.}}{2015}]%
        {BW2015Christoffel}
\bibfield{author}{\bibinfo{person}{Fabian Christoffel}, \bibinfo{person}{Bibek
  Paudel}, \bibinfo{person}{Chris Newell}, {and} \bibinfo{person}{Abraham
  Bernstein}.} \bibinfo{year}{2015}\natexlab{}.
\newblock \showarticletitle{Blockbusters and Wallflowers: Accurate, Diverse,
  and Scalable Recommendations with Random Walks}. In
  \bibinfo{booktitle}{\emph{Proceedings of the 9th ACM Conference on
  Recommender Systems}}. \bibinfo{publisher}{ACM}, \bibinfo{pages}{163–170}.
\newblock


\bibitem[\protect\citeauthoryear{Fan, Zhu, Han, Shi, Hu, Ma, and Li}{Fan
  et~al\mbox{.}}{2019b}]%
        {MEIRec2019Fan}
\bibfield{author}{\bibinfo{person}{Shaohua Fan}, \bibinfo{person}{Junxiong
  Zhu}, \bibinfo{person}{Xiaotian Han}, \bibinfo{person}{Chuan Shi},
  \bibinfo{person}{Linmei Hu}, \bibinfo{person}{Biyu Ma}, {and}
  \bibinfo{person}{Yongliang Li}.} \bibinfo{year}{2019}\natexlab{b}.
\newblock \showarticletitle{Metapath-Guided Heterogeneous Graph Neural Network
  for Intent Recommendation}. In \bibinfo{booktitle}{\emph{Proceedings of the
  25th ACM SIGKDD International Conference on Knowledge Discovery \& Data
  Mining}}. \bibinfo{publisher}{{ACM}}, \bibinfo{pages}{2478–2486}.
\newblock


\bibitem[\protect\citeauthoryear{Fan, Ma, Li, He, Zhao, Tang, and Yin}{Fan
  et~al\mbox{.}}{2019a}]%
        {fan2019social}
\bibfield{author}{\bibinfo{person}{Wenqi Fan}, \bibinfo{person}{Yao Ma},
  \bibinfo{person}{Qing Li}, \bibinfo{person}{Yuan He},
  \bibinfo{person}{Yihong~Eric Zhao}, \bibinfo{person}{Jiliang Tang}, {and}
  \bibinfo{person}{Dawei Yin}.} \bibinfo{year}{2019}\natexlab{a}.
\newblock \showarticletitle{Graph Neural Networks for Social Recommendation}.
  In \bibinfo{booktitle}{\emph{Proceedings of the 28th International Conference
  on World Wide Web}}. \bibinfo{publisher}{IW3C2}, \bibinfo{pages}{417--426}.
\newblock


\bibitem[\protect\citeauthoryear{Fouss, Pirotte, Renders, and Saerens}{Fouss
  et~al\mbox{.}}{2007}]%
        {fouss2007random}
\bibfield{author}{\bibinfo{person}{Fran{\c{c}}ois Fouss},
  \bibinfo{person}{Alain Pirotte}, \bibinfo{person}{Jean{-}Michel Renders},
  {and} \bibinfo{person}{Marco Saerens}.} \bibinfo{year}{2007}\natexlab{}.
\newblock \showarticletitle{Random-Walk Computation of Similarities between
  Nodes of a Graph with Application to Collaborative Recommendation}.
\newblock \bibinfo{journal}{\emph{{IEEE} Trans. Knowl. Data Eng.}}
  \bibinfo{volume}{19}, \bibinfo{number}{3} (\bibinfo{year}{2007}),
  \bibinfo{pages}{355--369}.
\newblock


\bibitem[\protect\citeauthoryear{Gori and Pucci}{Gori and Pucci}{2007}]%
        {ItemRank2007Gori}
\bibfield{author}{\bibinfo{person}{Marco Gori} {and} \bibinfo{person}{Augusto
  Pucci}.} \bibinfo{year}{2007}\natexlab{}.
\newblock \showarticletitle{ItemRank: A Random-Walk Based Scoring Algorithm for
  Recommender Engines}. In \bibinfo{booktitle}{\emph{Proceedings of the 20th
  International Joint Conference on Artifical Intelligence}}.
  \bibinfo{publisher}{Morgan Kaufmann Publishers Inc.},
  \bibinfo{pages}{2766–2771}.
\newblock


\bibitem[\protect\citeauthoryear{He and McAuley}{He and McAuley}{2016}]%
        {he2016VBPR}
\bibfield{author}{\bibinfo{person}{Ruining He} {and} \bibinfo{person}{Julian
  McAuley}.} \bibinfo{year}{2016}\natexlab{}.
\newblock \showarticletitle{VBPR: Visual Bayesian Personalized Ranking from
  Implicit Feedback}. In \bibinfo{booktitle}{\emph{Proceedings of the Thirtieth
  AAAI Conference on Artificial Intelligence}}. \bibinfo{publisher}{AAAI
  Press}, \bibinfo{pages}{144–150}.
\newblock


\bibitem[\protect\citeauthoryear{He, Chen, Kan, and Chen}{He
  et~al\mbox{.}}{2015}]%
        {he2015trirank}
\bibfield{author}{\bibinfo{person}{Xiangnan He}, \bibinfo{person}{Tao Chen},
  \bibinfo{person}{Min-Yen Kan}, {and} \bibinfo{person}{Xiao Chen}.}
  \bibinfo{year}{2015}\natexlab{}.
\newblock \showarticletitle{TriRank: Reviewaware Explainable Recommendation by
  Modeling Aspects}. In \bibinfo{booktitle}{\emph{Proceedings of the 24th {ACM}
  International Conference on Information and Knowledge Management}}.
  \bibinfo{publisher}{ACM}, \bibinfo{pages}{1661--1670}.
\newblock


\bibitem[\protect\citeauthoryear{He, Deng, Wang, Li, Zhang, and Wang}{He
  et~al\mbox{.}}{2020}]%
        {He2020lightgcn}
\bibfield{author}{\bibinfo{person}{Xiangnan He}, \bibinfo{person}{Kuan Deng},
  \bibinfo{person}{Xiang Wang}, \bibinfo{person}{Yan Li},
  \bibinfo{person}{YongDong Zhang}, {and} \bibinfo{person}{Meng Wang}.}
  \bibinfo{year}{2020}\natexlab{}.
\newblock \showarticletitle{LightGCN: Simplifying and Powering Graph
  Convolution Network for Recommendation}. In
  \bibinfo{booktitle}{\emph{Proceedings of the 43rd International ACM SIGIR
  Conference on Research and Development in Information Retrieval}}.
  \bibinfo{publisher}{ACM}, \bibinfo{pages}{639–648}.
\newblock


\bibitem[\protect\citeauthoryear{He, Liao, Zhang, Nie, Hu, and Chua}{He
  et~al\mbox{.}}{2017}]%
        {he2017neural}
\bibfield{author}{\bibinfo{person}{Xiangnan He}, \bibinfo{person}{Lizi Liao},
  \bibinfo{person}{Hanwang Zhang}, \bibinfo{person}{Liqiang Nie},
  \bibinfo{person}{Xia Hu}, {and} \bibinfo{person}{Tat-Seng Chua}.}
  \bibinfo{year}{2017}\natexlab{}.
\newblock \showarticletitle{Neural collaborative filtering}. In
  \bibinfo{booktitle}{\emph{Proceedings of the 26th International Conference on
  World Wide Web}}. \bibinfo{publisher}{IW3C2}, \bibinfo{pages}{173--182}.
\newblock


\bibitem[\protect\citeauthoryear{Hsieh, Yang, Cui, Lin, Belongie, and
  Estrin}{Hsieh et~al\mbox{.}}{2017}]%
        {hsieh2017cml}
\bibfield{author}{\bibinfo{person}{Cheng-Kang Hsieh}, \bibinfo{person}{Longqi
  Yang}, \bibinfo{person}{Yin Cui}, \bibinfo{person}{Tsung-Yi Lin},
  \bibinfo{person}{Serge Belongie}, {and} \bibinfo{person}{Deborah Estrin}.}
  \bibinfo{year}{2017}\natexlab{}.
\newblock \showarticletitle{Collaborative metric learning}. In
  \bibinfo{booktitle}{\emph{Proceedings of the 26th International Conference on
  World Wide Web}}. \bibinfo{publisher}{IW3C2}, \bibinfo{pages}{193--201}.
\newblock


\bibitem[\protect\citeauthoryear{Hu, Ren, Luo, Zhan, and Li}{Hu
  et~al\mbox{.}}{2019}]%
        {HuY2019}
\bibfield{author}{\bibinfo{person}{Yupeng Hu}, \bibinfo{person}{Pengjie Ren},
  \bibinfo{person}{Wei Luo}, \bibinfo{person}{Peng Zhan}, {and}
  \bibinfo{person}{Xueqing Li}.} \bibinfo{year}{2019}\natexlab{}.
\newblock \showarticletitle{Multi-resolution representation with recurrent
  neural networks application for streaming time series in IoT}.
\newblock \bibinfo{journal}{\emph{Computer Networks}}  \bibinfo{volume}{152}
  (\bibinfo{year}{2019}), \bibinfo{pages}{114--132}.
\newblock


\bibitem[\protect\citeauthoryear{Hu, Zhan, Xu, Zhao, Li, and Li}{Hu
  et~al\mbox{.}}{2020}]%
        {hu2020temporal}
\bibfield{author}{\bibinfo{person}{Yupeng Hu}, \bibinfo{person}{Peng Zhan},
  \bibinfo{person}{Yang Xu}, \bibinfo{person}{Jia Zhao}, \bibinfo{person}{Yujun
  Li}, {and} \bibinfo{person}{Xueqing Li}.} \bibinfo{year}{2020}\natexlab{}.
\newblock \showarticletitle{Temporal representation learning for time series
  classification}.
\newblock \bibinfo{journal}{\emph{Neural Computing and Applications}}
  (\bibinfo{year}{2020}), \bibinfo{pages}{1--14}.
\newblock


\bibitem[\protect\citeauthoryear{Kingma and Ba}{Kingma and Ba}{2015}]%
        {kingma2014adam}
\bibfield{author}{\bibinfo{person}{Diederik~P Kingma} {and}
  \bibinfo{person}{Jimmy Ba}.} \bibinfo{year}{2015}\natexlab{}.
\newblock \showarticletitle{Adam: A method for stochastic optimization}. In
  \bibinfo{booktitle}{\emph{Proceedings of the 3rd International Conference on
  Learning Representations}}.
\newblock


\bibitem[\protect\citeauthoryear{Kipf and Welling}{Kipf and Welling}{2017}]%
        {kipf2017gcn}
\bibfield{author}{\bibinfo{person}{Thomas~N. Kipf} {and} \bibinfo{person}{Max
  Welling}.} \bibinfo{year}{2017}\natexlab{}.
\newblock \showarticletitle{Semi-Supervised Classification with Graph
  Convolutional Networks}. In \bibinfo{booktitle}{\emph{ICLR}}.
  \bibinfo{publisher}{OpenReview.net}.
\newblock


\bibitem[\protect\citeauthoryear{Koren, Bell, and Volinsky}{Koren
  et~al\mbox{.}}{2009}]%
        {Koren2009MF}
\bibfield{author}{\bibinfo{person}{Yehuda Koren}, \bibinfo{person}{Robert
  Bell}, {and} \bibinfo{person}{Chris Volinsky}.}
  \bibinfo{year}{2009}\natexlab{}.
\newblock \showarticletitle{Matrix factorization techniques for recommender
  systems}. In \bibinfo{booktitle}{\emph{{IEEE} Computer}},
  Vol.~\bibinfo{volume}{42}. \bibinfo{pages}{42--49}.
\newblock


\bibitem[\protect\citeauthoryear{Liu, Cheng, Sun, Wang, Nie, and
  Kankanhalli}{Liu et~al\mbox{.}}{2019}]%
        {Liu2019MAML}
\bibfield{author}{\bibinfo{person}{Fan Liu}, \bibinfo{person}{Zhiyong Cheng},
  \bibinfo{person}{Changchang Sun}, \bibinfo{person}{Yinglong Wang},
  \bibinfo{person}{Liqiang Nie}, {and} \bibinfo{person}{Mohan Kankanhalli}.}
  \bibinfo{year}{2019}\natexlab{}.
\newblock \showarticletitle{User Diverse Preference Modeling by Multimodal
  Attentive Metric Learning}. In \bibinfo{booktitle}{\emph{Proceedings of the
  27th {ACM} International Conference on Multimedia}}.
  \bibinfo{publisher}{ACM}, \bibinfo{pages}{1526–1534}.
\newblock


\bibitem[\protect\citeauthoryear{Liu, Cheng, Zhu, Liu, and Nie}{Liu
  et~al\mbox{.}}{2020a}]%
        {liu2020A2GCN}
\bibfield{author}{\bibinfo{person}{Fan Liu}, \bibinfo{person}{Zhiyong Cheng},
  \bibinfo{person}{Lei Zhu}, \bibinfo{person}{Chenghao Liu}, {and}
  \bibinfo{person}{Liqiang Nie}.} \bibinfo{year}{2020}\natexlab{a}.
\newblock \showarticletitle{An Attribute-aware Attentive GCN Model for
  Recommendation}.
\newblock \bibinfo{journal}{\emph{IEEE Transactions on Knowledge and Data
  Engineering}} (\bibinfo{year}{2020}), \bibinfo{pages}{1--12}.
\newblock


\bibitem[\protect\citeauthoryear{Liu and Wu}{Liu and Wu}{2015}]%
        {Liu2015CACF}
\bibfield{author}{\bibinfo{person}{Xin Liu} {and} \bibinfo{person}{Wei Wu}.}
  \bibinfo{year}{2015}\natexlab{}.
\newblock \showarticletitle{Learning Context-Aware Latent Representations for
  Context-Aware Collaborative Filtering}. In
  \bibinfo{booktitle}{\emph{Proceedings of the 38th International ACM SIGIR
  Conference on Research and Development in Information Retrieval}}.
  \bibinfo{publisher}{Association for Computing Machinery},
  \bibinfo{pages}{887–890}.
\newblock


\bibitem[\protect\citeauthoryear{Liu, Ren, Zhang, Che, Liu, and Yin}{Liu
  et~al\mbox{.}}{2020b}]%
        {Liu2020KGIE}
\bibfield{author}{\bibinfo{person}{Yuanxing Liu}, \bibinfo{person}{Zhaochun
  Ren}, \bibinfo{person}{Wei-Nan Zhang}, \bibinfo{person}{Wanxiang Che},
  \bibinfo{person}{Ting Liu}, {and} \bibinfo{person}{Dawei Yin}.}
  \bibinfo{year}{2020}\natexlab{b}.
\newblock \showarticletitle{Keywords Generation Improves E-Commerce
  Session-Based Recommendation}. In \bibinfo{booktitle}{\emph{Proceedings of
  The Web Conference 2020}}. \bibinfo{publisher}{Association for Computing
  Machinery}, \bibinfo{pages}{1604–1614}.
\newblock


\bibitem[\protect\citeauthoryear{Maas, Hannun, and Ng}{Maas
  et~al\mbox{.}}{2013}]%
        {Andrew2013leaky}
\bibfield{author}{\bibinfo{person}{Andrew~L Maas}, \bibinfo{person}{Awni~Y
  Hannun}, {and} \bibinfo{person}{Andrew~Y Ng}.}
  \bibinfo{year}{2013}\natexlab{}.
\newblock \showarticletitle{Rectifier nonlinearities improve neural network
  acoustic models}. In \bibinfo{booktitle}{\emph{ICML Workshop on Deep Learning
  for Audio, Speech and Language Processing}}.
\newblock


\bibitem[\protect\citeauthoryear{McAuley and Leskovec}{McAuley and
  Leskovec}{2013}]%
        {mcauley2013hidden}
\bibfield{author}{\bibinfo{person}{Julian McAuley} {and} \bibinfo{person}{Jure
  Leskovec}.} \bibinfo{year}{2013}\natexlab{}.
\newblock \showarticletitle{Hidden factors and hidden topics: understanding
  rating dimensions with review text}. In \bibinfo{booktitle}{\emph{Proceedings
  of the 7th {ACM} Conference on Recommender Systems}}.
  \bibinfo{publisher}{ACM}, \bibinfo{pages}{165--172}.
\newblock


\bibitem[\protect\citeauthoryear{Qiu, Tang, Ma, Dong, Wang, and Tang}{Qiu
  et~al\mbox{.}}{2018}]%
        {qiu2018deepinf}
\bibfield{author}{\bibinfo{person}{Jiezhong Qiu}, \bibinfo{person}{Jian Tang},
  \bibinfo{person}{Hao Ma}, \bibinfo{person}{Yuxiao Dong},
  \bibinfo{person}{Kuansan Wang}, {and} \bibinfo{person}{Jie Tang}.}
  \bibinfo{year}{2018}\natexlab{}.
\newblock \showarticletitle{DeepInf: Social Influence Prediction with Deep
  Learning.}. In \bibinfo{booktitle}{\emph{Proceedings of the 24th {ACM}
  {SIGKDD} International Conference on Knowledge Discovery and Data Mining}}.
  \bibinfo{publisher}{ACM}, \bibinfo{pages}{2110--2119}.
\newblock


\bibitem[\protect\citeauthoryear{Rendle, Freudenthaler, Gantner, and
  Schmidt-Thieme}{Rendle et~al\mbox{.}}{2009}]%
        {rendle2009bpr}
\bibfield{author}{\bibinfo{person}{Steffen Rendle}, \bibinfo{person}{Christoph
  Freudenthaler}, \bibinfo{person}{Zeno Gantner}, {and} \bibinfo{person}{Lars
  Schmidt-Thieme}.} \bibinfo{year}{2009}\natexlab{}.
\newblock \showarticletitle{{BPR}: Bayesian personalized ranking from implicit
  feedback}. In \bibinfo{booktitle}{\emph{UAI}}. \bibinfo{pages}{452--461}.
\newblock


\bibitem[\protect\citeauthoryear{Sun, Zhang, Ma, Coates, Guo, Tang, and He}{Sun
  et~al\mbox{.}}{2019}]%
        {Sun2019MGCCF}
\bibfield{author}{\bibinfo{person}{Jianing Sun}, \bibinfo{person}{Yingxue
  Zhang}, \bibinfo{person}{Chen Ma}, \bibinfo{person}{Mark Coates},
  \bibinfo{person}{Huifeng Guo}, \bibinfo{person}{Ruiming Tang}, {and}
  \bibinfo{person}{Xiuqiang He}.} \bibinfo{year}{2019}\natexlab{}.
\newblock \showarticletitle{Multi-graph convolution collaborative filtering}.
  In \bibinfo{booktitle}{\emph{Proceedings of IEEE International Conference on
  Data Mining}}. \bibinfo{pages}{1306 -- 1311}.
\newblock


\bibitem[\protect\citeauthoryear{van~den Berg, Kipf, and Welling}{van~den Berg
  et~al\mbox{.}}{2018}]%
        {berg2019gcmc}
\bibfield{author}{\bibinfo{person}{Rianne van~den Berg},
  \bibinfo{person}{Thomas~N. Kipf}, {and} \bibinfo{person}{Max Welling}.}
  \bibinfo{year}{2018}\natexlab{}.
\newblock \showarticletitle{Graph Convolutional Matrix Completion}. In
  \bibinfo{booktitle}{\emph{ACM SIGKDD: Deep Learning Day}}.
  \bibinfo{publisher}{ACM}.
\newblock


\bibitem[\protect\citeauthoryear{Wang, Zhang, Wang, Zhao, Li, Xie, and
  Guo}{Wang et~al\mbox{.}}{2018}]%
        {wang2018ripplenet}
\bibfield{author}{\bibinfo{person}{Hongwei Wang}, \bibinfo{person}{Fuzheng
  Zhang}, \bibinfo{person}{Jialin Wang}, \bibinfo{person}{Miao Zhao},
  \bibinfo{person}{Wenjie Li}, \bibinfo{person}{Xing Xie}, {and}
  \bibinfo{person}{Minyi Guo}.} \bibinfo{year}{2018}\natexlab{}.
\newblock \showarticletitle{RippleNet: Propagating User Preferences on the
  Knowledge Graph for Recommender Systems}. In
  \bibinfo{booktitle}{\emph{CIKM}}. \bibinfo{publisher}{{ACM}},
  \bibinfo{pages}{417--426}.
\newblock


\bibitem[\protect\citeauthoryear{Wang, Zhang, Wang, Zhao, Li, Xie, and
  Guo}{Wang et~al\mbox{.}}{2019c}]%
        {wang2019tois}
\bibfield{author}{\bibinfo{person}{Hongwei Wang}, \bibinfo{person}{Fuzheng
  Zhang}, \bibinfo{person}{Jialin Wang}, \bibinfo{person}{Miao Zhao},
  \bibinfo{person}{Wenjie Li}, \bibinfo{person}{Xing Xie}, {and}
  \bibinfo{person}{Minyi Guo}.} \bibinfo{year}{2019}\natexlab{c}.
\newblock \showarticletitle{Exploring High-Order User Preference on the
  Knowledge Graph for Recommender Systems}.
\newblock \bibinfo{journal}{\emph{{ACM} Trans. Inf. Syst.}}
  \bibinfo{volume}{37}, \bibinfo{number}{3} (\bibinfo{year}{2019}),
  \bibinfo{pages}{32:1--32:26}.
\newblock


\bibitem[\protect\citeauthoryear{Wang, He, Cao, Liu, and Chua}{Wang
  et~al\mbox{.}}{2019a}]%
        {wang2019kdd}
\bibfield{author}{\bibinfo{person}{Xiang Wang}, \bibinfo{person}{Xiangnan He},
  \bibinfo{person}{Yixin Cao}, \bibinfo{person}{Meng Liu}, {and}
  \bibinfo{person}{Tat-Seng Chua}.} \bibinfo{year}{2019}\natexlab{a}.
\newblock \showarticletitle{{KGAT:} Knowledge Graph Attention Network for
  Recommendation}. In \bibinfo{booktitle}{\emph{Proceedings of the 25th {ACM}
  {SIGKDD} International Conference on Knowledge Discovery and Data Mining}}.
  \bibinfo{publisher}{ACM}, \bibinfo{pages}{950--958}.
\newblock


\bibitem[\protect\citeauthoryear{Wang, He, Wang, Feng, and Chua}{Wang
  et~al\mbox{.}}{2019b}]%
        {wang2019ngcf}
\bibfield{author}{\bibinfo{person}{Xiang Wang}, \bibinfo{person}{Xiangnan He},
  \bibinfo{person}{Meng Wang}, \bibinfo{person}{Fuli Feng}, {and}
  \bibinfo{person}{Tat-Seng Chua}.} \bibinfo{year}{2019}\natexlab{b}.
\newblock \showarticletitle{Neural Graph Collaborative Filtering}. In
  \bibinfo{booktitle}{\emph{Proceedings of the 42nd International {ACM} {SIGIR}
  Conference on Research and Development in Information Retrieval}}.
  \bibinfo{publisher}{ACM}, \bibinfo{pages}{165--174}.
\newblock


\bibitem[\protect\citeauthoryear{Wei, Cheng, Yu, Zhao, Zhu, and Nie}{Wei
  et~al\mbox{.}}{2019a}]%
        {wei2019mm}
\bibfield{author}{\bibinfo{person}{Yinwei Wei}, \bibinfo{person}{Zhiyong
  Cheng}, \bibinfo{person}{Xuzheng Yu}, \bibinfo{person}{Zhou Zhao},
  \bibinfo{person}{Lei Zhu}, {and} \bibinfo{person}{Liqiang Nie}.}
  \bibinfo{year}{2019}\natexlab{a}.
\newblock \showarticletitle{Personalized Hashtag Recommendation for
  Micro-videos}. In \bibinfo{booktitle}{\emph{Proceedings of the 27th {ACM}
  International Conference on Multimedia}}. \bibinfo{publisher}{ACM},
  \bibinfo{pages}{1446--1454}.
\newblock


\bibitem[\protect\citeauthoryear{Wei, Wang, Nie, He, and Chua}{Wei
  et~al\mbox{.}}{2020}]%
        {wei2020GRCN}
\bibfield{author}{\bibinfo{person}{Yinwei Wei}, \bibinfo{person}{Xiang Wang},
  \bibinfo{person}{Liqiang Nie}, \bibinfo{person}{Xiangnan He}, {and}
  \bibinfo{person}{Tat-Seng Chua}.} \bibinfo{year}{2020}\natexlab{}.
\newblock \showarticletitle{Graph-Refined Convolutional Network for Multimedia
  Recommendation with Implicit Feedback}. In
  \bibinfo{booktitle}{\emph{Proceedings of the 28th ACM International
  Conference on Multimedia}}. \bibinfo{publisher}{ACM},
  \bibinfo{pages}{3541--3549}.
\newblock


\bibitem[\protect\citeauthoryear{Wei, Wang, Nie, He, Hong, and Chua}{Wei
  et~al\mbox{.}}{2019b}]%
        {wei2019MMGCN}
\bibfield{author}{\bibinfo{person}{Yinwei Wei}, \bibinfo{person}{Xiang Wang},
  \bibinfo{person}{Liqiang Nie}, \bibinfo{person}{Xiangnan He},
  \bibinfo{person}{Richang Hong}, {and} \bibinfo{person}{Tat-Seng Chua}.}
  \bibinfo{year}{2019}\natexlab{b}.
\newblock \showarticletitle{MMGCN: Multi-modal graph convolution network for
  personalized recommendation of micro-video}. In
  \bibinfo{booktitle}{\emph{Proceedings of the 27th ACM International
  Conference on Multimedia}}. \bibinfo{publisher}{ACM},
  \bibinfo{pages}{1437--1445}.
\newblock


\bibitem[\protect\citeauthoryear{Wu, Souza, Zhang, Fifty, Yu, and
  Weinberger}{Wu et~al\mbox{.}}{2019}]%
        {pmlr-v97-wu19e}
\bibfield{author}{\bibinfo{person}{Felix Wu}, \bibinfo{person}{Amauri Souza},
  \bibinfo{person}{Tianyi Zhang}, \bibinfo{person}{Christopher Fifty},
  \bibinfo{person}{Tao Yu}, {and} \bibinfo{person}{Kilian Weinberger}.}
  \bibinfo{year}{2019}\natexlab{}.
\newblock \showarticletitle{Simplifying Graph Convolutional Networks}.
  \bibinfo{publisher}{PMLR}, \bibinfo{pages}{6861--6871}.
\newblock


\bibitem[\protect\citeauthoryear{Wu, DuBois, Zheng, and Ester}{Wu
  et~al\mbox{.}}{2016}]%
        {wu2016cdae}
\bibfield{author}{\bibinfo{person}{Yao Wu}, \bibinfo{person}{Christopher
  DuBois}, \bibinfo{person}{Alice~X. Zheng}, {and} \bibinfo{person}{Martin
  Ester}.} \bibinfo{year}{2016}\natexlab{}.
\newblock \showarticletitle{Collaborative Denoising Auto-Encoders for Top-N
  Recommender Systems}. In \bibinfo{booktitle}{\emph{Proceedings of the 9th
  {ACM} International Conference on Web Search and Data Mining}}.
  \bibinfo{publisher}{ACM}, \bibinfo{pages}{153--162}.
\newblock


\bibitem[\protect\citeauthoryear{Xavier and Yoshua}{Xavier and Yoshua}{2010}]%
        {Xavier2010xavier}
\bibfield{author}{\bibinfo{person}{Glorot Xavier} {and} \bibinfo{person}{Bengio
  Yoshua}.} \bibinfo{year}{2010}\natexlab{}.
\newblock \showarticletitle{Understanding the difficulty of training deep
  feedforward neural networks}. In \bibinfo{booktitle}{\emph{Proceedings of the
  13th International Conference on Artificial Intelligence and Statistics}}.
  \bibinfo{publisher}{JMLR}, \bibinfo{pages}{249--256}.
\newblock


\bibitem[\protect\citeauthoryear{Xin, Karatzoglou, Arapakis, and Jose}{Xin
  et~al\mbox{.}}{2020}]%
        {Xin2020GraphHN}
\bibfield{author}{\bibinfo{person}{Xin Xin}, \bibinfo{person}{Alexandros
  Karatzoglou}, \bibinfo{person}{I. Arapakis}, {and} \bibinfo{person}{J.
  Jose}.} \bibinfo{year}{2020}\natexlab{}.
\newblock \showarticletitle{Graph Highway Networks}.
\newblock \bibinfo{journal}{\emph{ArXiv}}  \bibinfo{volume}{abs/2004.04635}
  (\bibinfo{year}{2020}).
\newblock


\bibitem[\protect\citeauthoryear{Xue, Dai, Zhang, Huang, and Chen}{Xue
  et~al\mbox{.}}{2017}]%
        {xue2017deep}
\bibfield{author}{\bibinfo{person}{HongJian Xue}, \bibinfo{person}{XinYu Dai},
  \bibinfo{person}{Jianbing Zhang}, \bibinfo{person}{Shujian Huang}, {and}
  \bibinfo{person}{Jiajun Chen}.} \bibinfo{year}{2017}\natexlab{}.
\newblock \showarticletitle{Deep matrix factorization models for recommender
  systems}. In \bibinfo{booktitle}{\emph{Proceedings of the 26h International
  Joint Conference on Artificial Intelligence}}. \bibinfo{publisher}{AAAI
  Press}, \bibinfo{pages}{3203--3209}.
\newblock


\bibitem[\protect\citeauthoryear{Yang, Chen, Wang, and Tsai}{Yang
  et~al\mbox{.}}{2018}]%
        {hoprec}
\bibfield{author}{\bibinfo{person}{Jheng{-}Hong Yang},
  \bibinfo{person}{Chih{-}Ming Chen}, \bibinfo{person}{Chuan{-}Ju Wang}, {and}
  \bibinfo{person}{Ming{-}Feng Tsai}.} \bibinfo{year}{2018}\natexlab{}.
\newblock \showarticletitle{{HOP}-rec: high-order proximity for implicit
  recommendation}. In \bibinfo{booktitle}{\emph{RecSys}}.
  \bibinfo{publisher}{ACM}, \bibinfo{pages}{140--144}.
\newblock


\bibitem[\protect\citeauthoryear{Ying, He, Chen, Eksombatchai, Hamilton, and
  Leskovec}{Ying et~al\mbox{.}}{2018}]%
        {ying2018pinsage}
\bibfield{author}{\bibinfo{person}{Rex Ying}, \bibinfo{person}{Ruining He},
  \bibinfo{person}{Kaifeng Chen}, \bibinfo{person}{Pong Eksombatchai},
  \bibinfo{person}{William~L. Hamilton}, {and} \bibinfo{person}{Jure
  Leskovec}.} \bibinfo{year}{2018}\natexlab{}.
\newblock \showarticletitle{Graph Convolutional Neural Networks for Web-Scale
  Recommender Systems}. In \bibinfo{booktitle}{\emph{Proceedings of the 24th
  {ACM} {SIGKDD} International Conference on Knowledge Discovery and Data
  Mining}}. \bibinfo{publisher}{ACM}, \bibinfo{pages}{974--983}.
\newblock


\bibitem[\protect\citeauthoryear{Yu, Zhang, Pei, Sun, and Zhang}{Yu
  et~al\mbox{.}}{2018}]%
        {yu2018walkranker}
\bibfield{author}{\bibinfo{person}{Lu Yu}, \bibinfo{person}{Chuxu Zhang},
  \bibinfo{person}{Shichao Pei}, \bibinfo{person}{Guolei Sun}, {and}
  \bibinfo{person}{Xiangliang Zhang}.} \bibinfo{year}{2018}\natexlab{}.
\newblock \showarticletitle{WalkRanker: {A} Unified Pairwise Ranking Model With
  Multiple Relations for Item Recommendation}. In
  \bibinfo{booktitle}{\emph{IJCAI}}. \bibinfo{publisher}{{AAAI} Press},
  \bibinfo{pages}{2596--2603}.
\newblock


\bibitem[\protect\citeauthoryear{Zhang and McAuley}{Zhang and McAuley}{2020}]%
        {Zhang2020SMOGCF}
\bibfield{author}{\bibinfo{person}{Hengrui Zhang} {and} \bibinfo{person}{Julian
  McAuley}.} \bibinfo{year}{2020}\natexlab{}.
\newblock \showarticletitle{Stacked Mixed-Order Graph Convolutional Networks
  for Collaborative Filtering}. In \bibinfo{booktitle}{\emph{Proceedings of the
  2020 SIAM International Conference on Data Mining}}.
  \bibinfo{publisher}{Society for Industrial and Applied Mathematics},
  \bibinfo{pages}{73--81}.
\newblock


\end{thebibliography}
\end{document}